\documentclass[doublecol]{epl2}

\usepackage{dcolumn}
\usepackage{bm}
\usepackage{graphicx,color}
\usepackage{amsmath}
\usepackage{amssymb}
\usepackage{verbatim}
\usepackage{subfigure}

\newcommand\Dfrtl[1]{\ensuremath{\,\mathrm{d}#1\,}}
\newcommand\imag{\ensuremath{\mathrm{i}}}

\newcommand\euler[1]{\ensuremath{\mathrm{e}^{#1}}}


\abstract{
We use different numerical approaches to calculate 
the double occupancy and
magnetic susceptibility as a function of a bias voltage in an Anderson
impurity model. 
Specifically, we compare results from the Matsubara-voltage
quantum Monte-Carlo approach (MV-QMC),
the scattering-states numerical renormalization group (SNRG), and real-time
quantum Monte-Carlo (RT-QMC), covering Coulomb repulsions ranging
from the weak-coupling well into the strong-coupling regime. 
We observe a distinctly different behavior of the double occupancy
and the magnetic response.
The former  measures charge fluctuations and thus only indirectly exhibits
the Kondo scale, while the latter exhibits structures on the scale of the equilibrium Kondo temperature.
The Matsubara-voltage approach and the scattering-states
numerical renormalization group yield consistent values for the
magnetic susceptibility in the Kondo limit. 
On the other hand, all three numerical methods
produce different results for the behavior of charge fluctuations in strongly interacting dots out of
equilibrium.
}

\author{A. Dirks\inst{1,2,3} \and S. Schmitt\inst{4}\footnote{Present address: Honda Research Institute GmbH, D-63073 Offenbach, Germany} \and J.E. Han\inst{2} \and F. Anders\inst{4} \and P. Werner\inst{5} \and T. Pruschke\inst{1}}
\shortauthor{A. Dirks \etal}

\institute{                    
  \inst{1} Institut f\"ur Theoretische Physik, Universit\"at G\"ottingen, D-37077 G\"ottingen, Germany \\
  \inst{2} Department of Physics, State University of New York at Buffalo, Buffalo, NY 14260, USA \\
  \inst{3} Department of Physics, Georgetown University, Washington, DC 20057, USA \\
  \inst{4} Lehrstuhl f\"ur Theoretische Physik II, Technische Universit\"at Dortmund, D-44221 Dortmund, Germany \\
  \inst{5} Department of Physics, University of Fribourg, CH-1700 Fribourg, Switzerland 
}

\title{
Double Occupancy and Magnetic Susceptibility of the Anderson Impurity Model out of Equilibrium
}

\date{\today}

\begin{document}

\maketitle

\section{Introduction}
The advances in nanostructuring of heterogeneous semiconductors and in the handling of  molecules 
have made it possible to  reproducibly build carefully designed nanometer-scale devices.
These generically consist of a few locally interacting degrees of freedom, for example   
in a quantum dot in contact with macroscopic leads. The spatial
confinement of the quantum dot electrons to a few nanometers implies a 
small electrical capacitance $C$ and, hence, a sizable  charging energy $U=e^2/C$.
The attraction of such devices stems from their  highly controllable properties \cite{KastnerSET1992,Yoffe2001}.
Substantially  increasing the coupling to the leads but 
still maintaining  the charge fluctuation scale below charging energy 
tunes the quantum dots into the experimentally accessible Kondo-regime \cite{NatureGoldhaberGordon1998, pustilnik,cronenwett, potok}. This regime is
characterized by the lifting of the Coulomb blockade \cite{KastnerSET1992} in the quantum transport
at temperatures below a dynamically generated small energy scale, called Kondo temperature $T_K$, which ubiquitously  shows up in physical properties \cite{hewson}.

The experiments performed on mesoscopic systems typically are measurements of transport properties
in the presence of external fields and voltage bias, making a theoretical description in terms of 
non-equilibrium statistical physics mandatory. 
The theoretical challenge of the Kondo regime is related to the change of ground state \cite{hewson}
upon cooling the system from an intermediate-temperature local-moment regime to the low-temperature regime
which manifests itself in the lifting of the Coulomb blockade at zero bias. 
This crossover cannot be reliably
accessed by any finite order perturbation theory in the Coulomb repulsion and
requires more sophisticated  analytical methods such as Bethe ansatz\cite{BetheAnsatz}
or numerical methods such as Wilsons numerical renormalization group (NRG)\cite{wilsonNRG75}. 

However, no exact solution for the transport properties through quantum dots at finite bias
exists for models of  interacting quantum many-body systems out of equilibrium.
Over the past two decades, several approaches have been developed to approximately or numerically solve such models, 
ranging from perturbation theory \cite{hershfield,meir,ueda} and renormalization approaches
\cite{schoeller,rosch,gezzi} to various numerical techniques
\cite{thorwart,werner_rtqmc,werner_rtqmc2,anders_snrg,dmrg,segal,cohen}.
Most of these approaches work well in certain limits, where Kondo physics is either not yet relevant or
-- due to strong external fields -- already suppressed. Accessing the crossover 
regime where external fields, in particular the bias voltage across the dot,  
are of the order of the Kondo temperature, remains a great challenge.

In this paper we present results for static quantities of a quantum dot in steady-state non equilibrium for
dot parameters, temperatures, and voltages that fall precisely into this
challenging regime. One approach employed here has been
recently proposed by Han and Heary \cite{han_method}. It maps the steady-state
non-equilibrium system onto an infinite set of auxiliary equilibrium statistical-physics problems. 
The latter are solved by a continuous-time quantum Monte-Carlo algorithm
\cite{gull_review, dirks2010}. The main 
challenge in this approach is to map the auxiliary systems back onto the real
one, which can be accomplished by a standard 
maximum entropy analytical continuation procedure \cite{mark_gubernatis}.
Details of the numerical procedure have been provided in a recent publication
\cite{paper1}. Here, we compare the results of this Matsubara-voltage
quantum Monte Carlo (MV-QMC) approach to data obtained with a scattering-states numerical renormalization group (SNRG) method 
\cite{anders_snrg, schmitt10,schmitt11} and real-time quantum Monte Carlo (RT-QMC) \cite{werner_rtqmc, werner_rtqmc2}.

\section{Model and Methods}
The simplest and most frequently used model for a quantum dot is the single-impurity Anderson model \cite{pustilnik}. In this model, the dot is described by a single molecular orbital, which
can accommodate up to two electrons. In the doubly occupied case  the Coulomb repulsion leads to a charging  energy of $U$.
The dot orbital is connected via single-particle tunneling of amplitude $t_\alpha$ 
to two continuous sets of non-interacting fermionic
baths, which are called \emph{source} and \emph{drain lead}, 
and are indicated by an index $\alpha=\pm1$. The leads can have different chemical potentials,
and the difference $e\Phi=\mu_{-1}-\mu_{+1}$ in chemical potentials 
represents the physical \emph{bias voltage}. With these conventions,
the Hamiltonian reads 
\begin{equation}
\begin{split}
H = & \sum_{\alpha k \sigma} \epsilon_{\alpha k\sigma} 
c^\dagger_{\alpha k} c_{\alpha k \sigma} + 
 \sum_{\sigma=\pm1} \left(
 \epsilon_d + \sigma B \right)
  d_\sigma^\dagger d_\sigma 
\\
&
+ U n_{d,\uparrow} n_{d,\downarrow} + \sum_{\alpha k \sigma} \left( \frac{t_{\alpha}}{\sqrt{\Omega}} c^\dagger_{\alpha k \sigma} d_\sigma +
\text{h.c.}\right),
\end{split}
\end{equation}
where $c^\dagger_{\alpha k \sigma}$ and $d_\sigma^\dagger $ are the usual fermionic creation
operators of electrons with spin $\sigma=\{\pm1\}=\{\uparrow,\downarrow\}$, 
in lead $\alpha=\pm1$ with momentum $k$
or on the dot ($\hbar=c=k_\mathrm{B}=1$).  
The corresponding single-particle energies are  $\epsilon_{\alpha k \sigma}$ and
$\epsilon_d$, respectively, and we added
the Zeeman energy $\sigma B =\sigma g \mu_{\rm B} H/2$ 
into the single-particle energy on the dot, which includes the effect of an external magnetic field $H$.
We will consider the wide-band limit for the leads with symmetric coupling $t_-=t_+=t$ and 
the particle-hole symmetric point $\epsilon_d = -U/2$. $\Omega$ denotes the
phase space volume.  As unit of energy we use the
Anderson width $\Gamma=2\pi t^2 {\cal N}_F$, where ${\cal N}_F$ is the conduction electron
density of states at the Fermi energy.

The method developed by Han and Heary employs a complexification of the
physical bias voltage $\Phi\to\imag\varphi_m$
with $\varphi_m=4\pi m/\beta$ (Matsubara voltages) \cite{han_method}, which results in an infinite set of auxiliary
equilibrium statistical-mechanics systems one can efficiently solve by state-of-the-art Monte Carlo
algorithms \cite{gull_review,dirks2010}. The mapping back to non-equilibrium quantities is done via
analytical continuation $\imag\varphi_m \to \Phi \pm \imag\delta$ from the Matsubara voltages to the real voltage.
A central goal is of course the calculation of transport properties. Unfortunately, the non-locality of the current operator 
and the rather complicated analytical 
structure of the local Green's functions render such calculations very difficult \cite{paper2}.
However, obtaining results for static local quantities, such as the double occupancy on
the dot or the magnetization, is relatively straightforward \cite{paper1}.

The full derivation of the formulas connecting those static quantities in the Matsubara voltage space
with the real-voltage expectation value were presented in Ref.~\cite{paper1}. 
For local observables $\mathcal{O}$, such as the double occupancy and magnetization of the dot,
one finds the representation 
\begin{equation}
\langle\mathcal{O}\rangle(\imag\varphi_m) =
\left.\langle\mathcal{O}\rangle\right|_{\varphi_m\to\infty} +
\int \frac{\varrho_\mathcal{O}(\varphi)}{(\imag\varphi_m - \Phi) - \varphi} \Dfrtl\varphi,
\label{eq:representation}
\end{equation}
with $\varrho_\mathcal{O}(\varphi)$ the spectral function.
The physical expectation value is then given by \cite{paper1}
\begin{equation}
\langle \mathcal{O} \rangle_\text{neq} = 
\left.\langle\mathcal{O}\rangle\right|_{\varphi_m\to\infty} -
\mathcal{P}\!\!\!\int
\frac{\varrho_\mathcal{O}(\varphi)}{\varphi} \Dfrtl\varphi\;\;,
\label{eq:physicalestimate}
\end{equation}
where $\mathcal{P}\!\!\int\cdots$ denotes the principal value integral.

Highly precise data for $\langle\mathcal{O}\rangle(\imag\varphi_m)$ can be obtained
even for large values of $\varphi_m$
from the effective equilibrium systems of the Matsubara voltage
representation by use of the continuous-time quantum Monte Carlo (CT-QMC) technique
\cite{dirks2010}. In contrast to equilibrium QMC for the Anderson impurity model,
we expect a sign- respectively phase-problem here due to the presence of a complex
quantity in the effective action. It turns out, however, that this phase problem is rather weak for small
to intermediate bias, although it can become significant for large bias voltage and very large
$\imag\varphi_m$.\footnote{For example, for
$\imag\varphi_{m=60}$ we find for certain model parameters ($U=8\Gamma$, $\beta\Gamma = 20.0$, $B=0.04\Gamma$) 
average phase factors $|\langle \euler{i
\gamma} \rangle| \approx 0.073$ at $e\Phi=2\Gamma$, and $|\langle \euler{i \gamma}
\rangle| \approx 0.40$ at $e\Phi=1\Gamma$.}

\begin{figure}
\includegraphics[width=\linewidth]{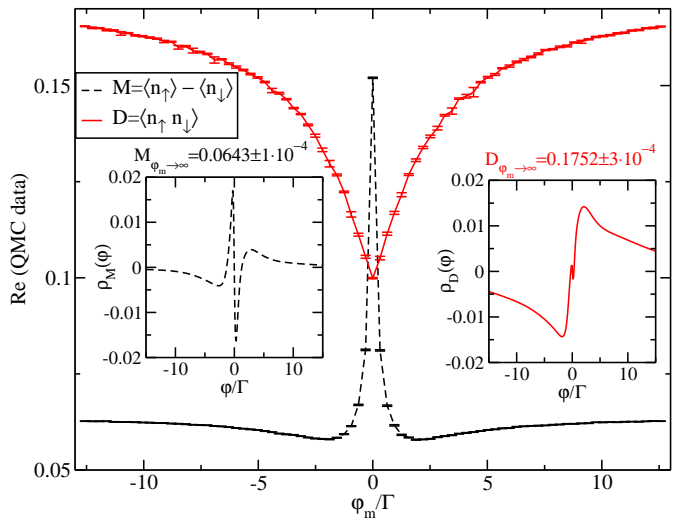}
\caption{Raw data used in the computation of the magnetization $M$ and double occupancy $D$ of a quantum dot ($U=8\Gamma$, $\beta\Gamma =40$, 
$B=0.02$, $e\Phi = 0.01\Gamma$) in
MV-QMC. The insets show the 
inferred offsets and spectral functions which yield the physical value.}
\label{fig:rawdataQMC}
\end{figure}

The representation \eqref{eq:representation} is formally similar to the
standard Lehmann-type representation of correlation functions. 
Due to the singular nature of the integral equation \eqref{eq:representation}, the numerical determination
of the spectral function $\varrho_\mathcal{O}(\varphi)$  from QMC data is known
to be an ill-posed problem. An adequate tool which helps to reduce
uncontrollable biases in the estimates of $\varrho_\mathcal{O}(\varphi)$ is
provided by the maximum entropy method (MaxEnt) \cite{mark_gubernatis}. It uses Bayesian inference by 
interpreting the spectral function as a probability density.
A-priori and a-posteriori information about this density and
quantities derived from it can be discriminated
and discussed properly.
As an illustration of the procedure, we plot in Fig.~\ref{fig:rawdataQMC}
the raw data for the calculation of the magnetization $M$ and double occupancy $D$ in a model with $U=8\Gamma$, $\beta\Gamma =40$, 
$B=0.02\Gamma$, $e\Phi = 0.01\Gamma$. The offsets inferred from these QMC results are $M_{i\varphi_n\rightarrow \infty}=0.0643(1)$ and $D_{i\varphi_n\rightarrow \infty}=0.1752(3)$ and the spectral functions obtained from the MaxEnt procedure are shown as insets. With these results, Eq.~(\ref{eq:physicalestimate}) leads to $M=0.134(1)$ and $D=0.107(9)$.

In order to analyze the numerical results of the Matsubara-voltage
technique in the strongly correlated regime quantitatively, we 
compare them to SNRG \cite{anders_snrg, schmitt10,schmitt11}  and real RT-QMC \cite{werner_rtqmc, werner_rtqmc2} 
 calculations. 

The SNRG method extends the well-known equilibrium numerical renormalization group
(NRG) \cite{wilsonNRG75}
to describe steady-state non-equilibrium transport through interacting
nano-devices. 
The methods starts from a formulation of the noninteracting $U=0$ problem 
in terms of exactly known scattering states, which are the solutions
of the Lippmann-Schwinger equations \cite{schmitt10}.
Therefore, the boundary conditions of
the open quantum system with particles entering and leaving the whole 
system are correctly incorporated. Formulating the NRG in the basis
of these scattering states allows for the application of the 
time-dependent NRG \cite{schiller2005,schiller2006}. Starting from the initial 
$U=0$ Hamiltonian the interaction $U$ is switched on 
and the time-evolution of the system in calculated.
The results for the long-time steady state limit can be 
accessed analytically and steady-state nonequilibrium 
expectation values are calculated for arbitrary interaction strength
and bias voltage.

The RT-QMC technique \cite{werner_rtqmc, werner_rtqmc2} is used to check
the  non-equilibrium double occupancies. This method computes
steady-state expectation values by performing a quench, either in the interaction $U$ or the voltage bias $\Phi$, and is based on a stochastic sampling of weak-coupling diagrams within the Keldysh real-time Green's function approach. 
The technique employed here is the interaction quench, described in detail in
Ref.~\cite{werner_rtqmc2}. We start from the noninteracting system
with applied bias voltage and switch on the interaction at time $t=0$. The
time-evolution of the double occupancy is then computed by randomly placing
interaction vertices on the Keldysh contour, using a Monte Carlo technique.
As a result of the quench, the double occupancy decreases and eventually
approaches a time-independent value corresponding to the steady-state double
occupancy of the interacting system \cite{werner_rtqmc}. If the times accessible 
in the RT-QMC calculation are long enough to see this convergence, the results are 
numerically exact.
Difficulties arise in the small-voltage regime, where the transient 
dynamics becomes slow (the longest accessible time is limited by a dynamical sign problem 
in the Monte Carlo sampling). The double occupancy is easier to measure than the magnetic susceptibility, because in a half-filled dot with symmetric bias, only even perturbation orders contribute to the observable, which reduces the dynamical sign problem.

\section{Results}

\subsection{Double occupancy}
\begin{figure}
\includegraphics[width=\linewidth]{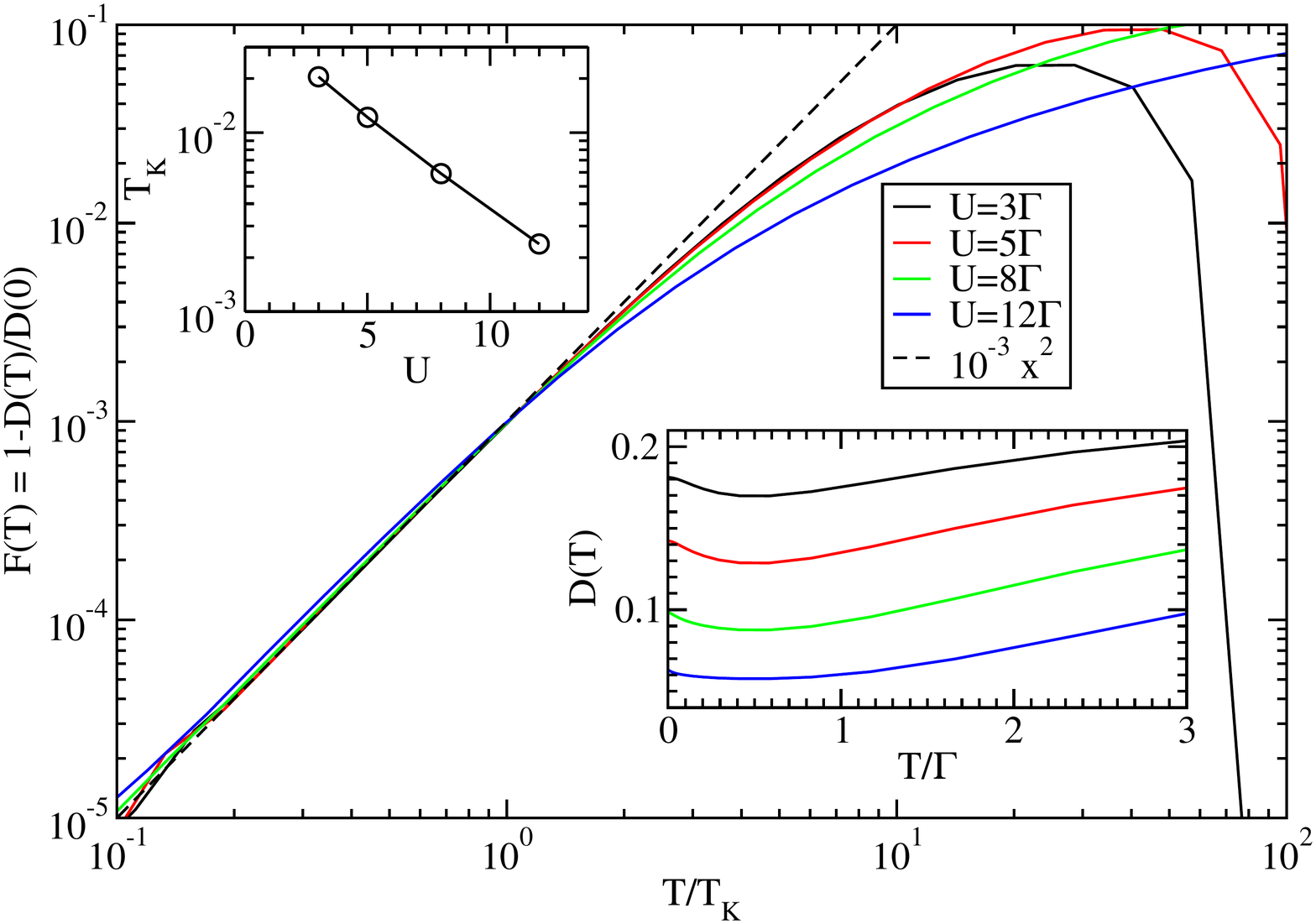}
\includegraphics[width=\linewidth]{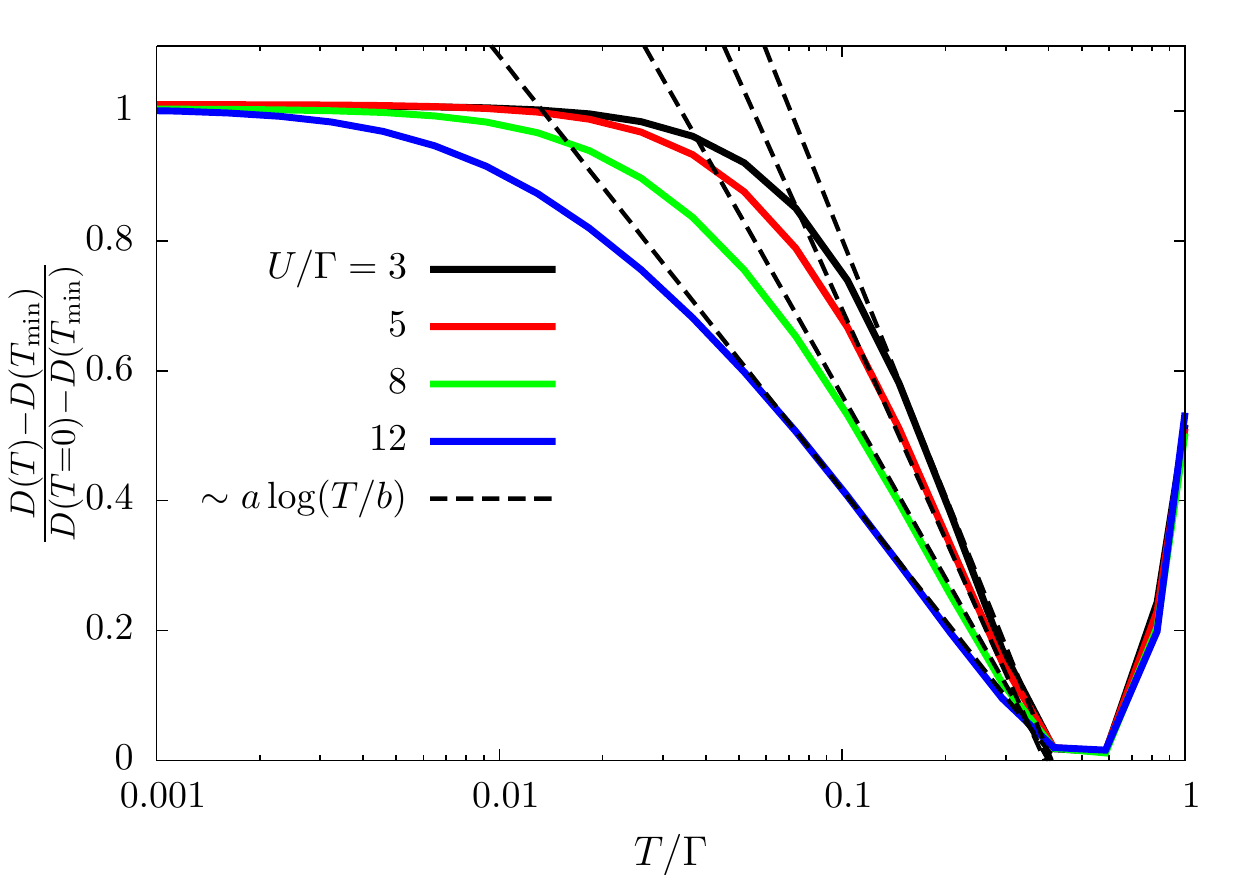}
\caption{ 
Equilibrium NRG results.
(a) The scaling function $F$ of the double occupancy $D$ as a function of temperature in units of the Kondo temperature
for various values of $U$. The dashed line shows the quadratic function $0.001 x^2$.
The lower right inset shows the double occupancy $D(T)$ as a function of temperature
in units of $\Gamma$.    
The upper left inset  shows the Kondo scale as extracted from a fit to the scaling function 
$F\approx 0.001 (T/T_K)^2$  as a function $U$.
(b) 
The increase of the double occupancy relative to its value
at the minimum, normalized to the total increase,  $\frac{D(T)-D(T_\mathrm{min})}{D(T=0)-D(T_\mathrm{min})}$
($T_\mathrm{min}$ is the temperature of the minimum). 
}
\label{fig:DoubleEqui}
\end{figure}

The double occupancy  $D = \langle n_\uparrow n_\downarrow \rangle$ 
is usually not discussed in the context of Kondo
physics, and therefore we believe it is useful to start here with 
a study of the equilibrium  behavior of this quantity
 in the absence of an external magnetic field, $B=0$. 
We consider the
particle-hole symmetric case  where
$\langle n_d \rangle = \langle n_\uparrow + n_\downarrow\rangle = 1$,
and the double occupancy 
measures the charge fluctuation of the quantum dot, i.e.\
\begin{equation}
D = \langle n_\uparrow n_\downarrow \rangle = 
\frac{1}{2} \langle (n_d - 1)^2 \rangle = \frac12\langle (n_d - \langle n_d \rangle)^2 \rangle.
\end{equation}
This quantity is shown as a function of $T$ in Fig.\ \ref{fig:DoubleEqui} for various values 
of the Coulomb repulsion $U$.
The high temperature limit  $D(T\gg\Gamma,U) \approx
\langle n_\uparrow \rangle \langle n_\downarrow\rangle = 1/4$ is approached by all
curves (not explicitly shown), 
and the low temperature saturation values roughly scale as $1/U$, as  expected
(see lower right inset of Fig.~\ref{fig:DoubleEqui}(a)).
However, the first striking observation is that, in contrast to e.g.\ the magnetization, 
one does not see any explicit signature of the Kondo scale. 
Instead, for all values of U a minimum appears for temperatures
on the scale of $T \approx \Gamma$ with
weakly temperature dependent tails at lower $T$.
The information about $T_\text{K}$ is contained in these tails. 
Plotting $F(T):=1-D(T)/D(T=0)$ as a function
of $T/T_\text{K}$ (main panel of Fig.\ \ref{fig:DoubleEqui}(a)) we find
a nice scaling for $T\lesssim T_\text{K}$. 
The curves for all $U$  fall on top of each other and follow 
a quadratic behavior, $F(T\to0)\sim (T/T_\text{K})^2$, which
is consistent with  the Fermi liquid  nature of the strong-coupling Kondo
fixed point at $T=0$. 
The Kondo scale, estimated
from the onset of this scaling behavior
via $F(T_K)\overset{!}= 0.001$, exhibits the expected exponential decrease 
as a function of $U$ (see upper left inset).

This unusual behavior 
must be interpreted in the following way. 
For higher $T$, as $U$ is increased, charge fluctuations are strongly suppressed by the Coulomb repulsion
and are frozen out
as the system approaches its local-moment fixed point. 
This happens on a scale
$\Gamma$, explaining why $D(T)$  flattens on that temperature scale for all
$U$. At very low 
temperatures, the systems tends to form a Kondo singlet
due to spin fluctuations, which are accompanied by virtual charge fluctuations, 
thereby again enhancing $D(T)$. (Note that the absolute value of this
contribution to $D(T)$ is of course again strongly suppressed with 
increasing $U$.)
This increase  produced by Kondo correlations  should therefore occur at temperatures
on the order of $T_\text{K}$. However, the enhancement of  $D(T)$  already
set in for temperatures
slightly below $\Gamma\gg T_\text{K}$ which is observable  in the lower right 
inset of Fig.~\ref{fig:DoubleEqui}(a)  and Fig.~\ref{fig:DoubleEqui}(b). 
This is due to the typical 
logarithmic tails\cite{tails} ubiquitous in Kondo systems as visible in Fig.~\ref{fig:DoubleEqui}(b).

\begin{figure}
\includegraphics[width=\linewidth]{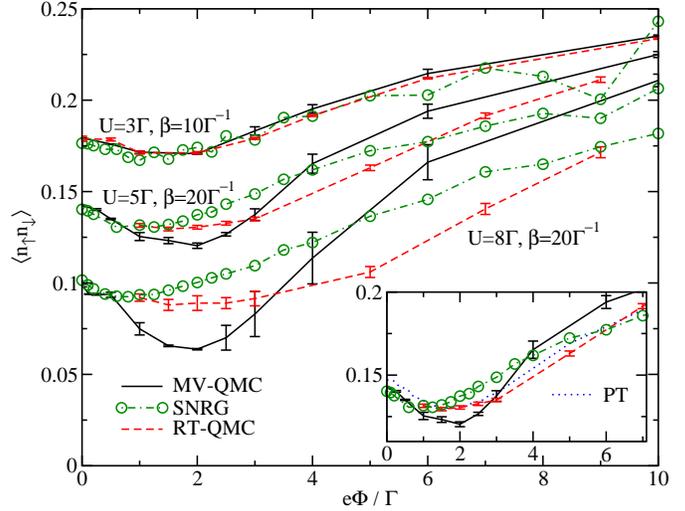}
\caption{(color online) Double occupancy as a function of the bias voltage for a broad range of the bias voltage.
The inset shows the double occupancy for $U=5\Gamma$, comparing to second-order
perturbation theory (PT). At the present stage of development, significant
mutual discrepancies are observed between all computational methods in the
intermediate voltage range.
}
\label{fig:docccompdiffU}
\end{figure}

In Fig.\ \ref{fig:docccompdiffU} we provide non-equilibrium double occupancy
data as a function of the bias voltage as obtained from the three approaches discussed above.  
The overall functional forms look quite similar to the equilibrium curves
of Fig.~\ref{fig:DoubleEqui}.
For large bias voltages all
methods approach the non-interacting
limit $\langle n_\uparrow n_\downarrow\rangle
\stackrel{\Phi\to\infty}{\longrightarrow} 
\langle n_\uparrow \rangle \langle n_\downarrow\rangle = 1/4$
and for zero bias voltage $\Phi=0$ they all reproduce the equilibrium value.
(There are no RT-QMC data available for
very small voltages, but it seems that the available data 
extrapolate to the SNRG and MV-QMC result.)
At intermediate voltages, all methods produce a minimum. For the weakest interaction, 
$U=3\Gamma$,  all curves agree quantitatively. However, 
for larger $U$, the approaches differ both in the position
and in the depth of this minimum. 

With increasing $U$, the MV-QMC approach exhibits the most pronounced minimum, whose
position slightly moves to larger voltages. The RT-QMC results show a more
shallow minimum, with a position roughly in agreement with MV-QMC. In contrast,
the depth of the minimum in the SNRG results does not deepen significantly with $U$, 
while the position even appears to move slightly towards \emph{smaller} 
voltages.

In Ref.~\cite{werner_rtqmc2} it was shown that low order perturbation theory 
results for the current are accurate up to $U/\Gamma\approx 4$, with small 
deviations in the
intermediate voltage regime for $U/\Gamma=6$. 
At larger
interactions, the perturbation theory result becomes incompatible
with RT-QMC in the intermediate bias regime $1 \lesssim V/\Gamma \lesssim 5$
\cite{werner_rtqmc2}. 
This is consistent with the findings of this study
as shown in the inset of Fig.~\ref{fig:docccompdiffU}, where the RT-QMC
results for $U/\Gamma$ are close to the prediction from second-order perturbation theory. 

The origin of the  discrepancies in the nonequilibrium double occupancy for intermediate and large $U$ is unclear.
If we assume that SNRG captures the correct position of the minimum in $D(\Phi)$, this would support 
the notion that a finite bias voltage can act similarly to an increase in 
temperature. The primary effect of both is to induce larger 
fluctuations, and therefore the  behavior
of the double occupancy as a function of temperature and bias voltage 
is qualitatively the same. Consequently, the position of the minimum 
does not move significantly with bias voltage and temperature.

On the other hand, the MV-QMC and RT-QMC results suggest that voltage and temperature 
act in a fundamentally different way on the double occupancy. The shift of the minimum position to larger voltages with increasing $U$
may be a genuine non-equilibrium effect, whose physical mechanism however remains to be clarified.

\subsection{Magnetic susceptibility}
\begin{figure}
\includegraphics[width=\linewidth]{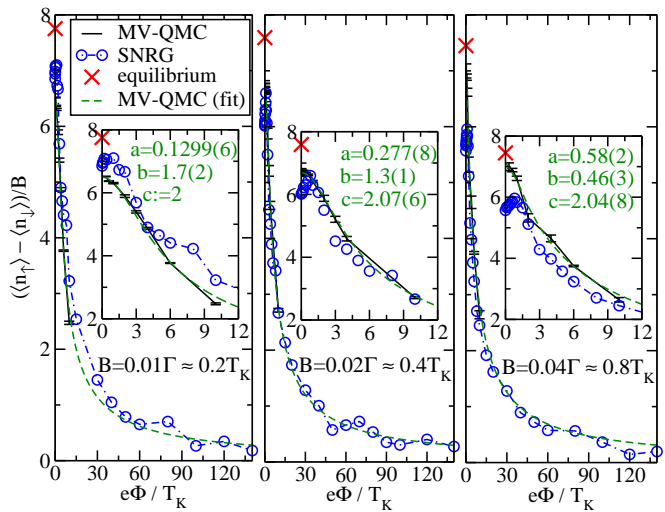}
\caption{
(Color online) Magnetic susceptibility as a function of the bias voltage
(in units of the Kondo temperature) obtained from MV-QMC and SNRG 
for $U=8\Gamma$, $\beta=40\Gamma^{-1}$ at different values of the magnetic field.
Within MV-QMC the computational effort is chosen to be identical for each of the curves which renders $B=0.01
\Gamma$ the least and $B=0.04\Gamma$ the most accurate.
The equilibrium limit is shown as the red cross on the $\Phi=0$ axis. The green dashed line shows 
a weighted least-square fit of the MV-QMC data to Eq.~\eqref{eq:fitmagn}.
We used the Haldane estimate of the equilibrium Kondo temperature 
$T_\mathrm{K} \approx \frac{1}{20} \Gamma$.
The insets show close-ups of the low voltage region, as well as parameters of the fit to Eq.\ \ref{eq:fitmagn}.
}
\label{fig:vglSNRGSuscept}
\end{figure}

In Fig.~\ref{fig:vglSNRGSuscept},
we present a comparison of the  SNRG and MV-QMC non-equilibrium data for the magnetic susceptibility
as a function of the bias voltage for three magnetic fields $B< T_K$.
The linear slope of the magnetization curves
in a small, but finite magnetic field $B$,
$\chi\sim \frac{\langle n_\uparrow\rangle-\langle n_\downarrow\rangle}B$,
is taken as an estimate for the susceptibility.
Note that the computational effort for the MV-QMC calculations is kept constant
throughout the considered parameter range, leading to a doubling  
of the relative statistical error as the magnetic field is 
decreased by a factor of two.

The magnetic susceptibility clearly exhibits a signature of the 
equilibrium Kondo scale, which is in contrast to the double occupancy. 
The susceptibility drops strongly when the voltage reaches the order 
of $T_\mathrm{K}$ indicating the destruction of the Kondo correlations
due to the voltage induced fluctuations. 
As a function of temperature, this is a well-known fact in equilibrium \cite{hewson}, and can also be shown to hold for
non-equilibrium systems, for example,  using a simple scaling analysis of simulation data
within the Matsubara-voltage formalism \cite{paper1}.

MV-QMC and SNRG agree reasonably well in the low-voltage regime. In both methods well-understood 
numerical issues lead to a discrepancy with the equilibrium value of the magnetization at
$\Phi =0$. In MV-QMC, this is due to a systematic bias in the MaxEnt 
estimator and the growth in statistical error at low fields \cite{paper1}. 
Within SNRG,  the the main source of inaccuracies is the truncation of the basis states due to the limited 
computer memory. Keeping the number of 
states fixed, this truncation becomes more severe in cases where many states become 
close in energy, which is the case when the voltage, the Kondo scale
and the magnetic field are of the same order.
A similar behavior of few-states SNRG can be observed for the
equilibrium temperature
dependence of $m(T)$ (not shown) which can, however, be overcome computationally
much more easily than in the non-equilibrium situation.

While the sign problem
in quantum Monte-Carlo prevents a direct computation of MV-QMC values at higher
voltages, this range is accessible with SNRG.
In Ref.~\cite{paper1} we noted that the Matsubara-voltage magnetization data can be described
by the expression
\begin{equation}
\frac{m(\tilde \Phi)}{B} \approx \frac{a}{B}\cdot \frac{1}{\frac{{\tilde
\Phi}^2}{\sqrt{b^2 + {\tilde\Phi}^2}} + c },
\label{eq:fitmagn}
\end{equation}
where $\tilde \Phi = \Phi /(\pi T_\mathrm{K})$. 
The additional square-root in the denominator of Eq.~\eqref{eq:fitmagn} 
leads to larger tails in the high-voltage regime of the susceptibility when compared
to a fit with a pure Lorentz-like function, $\sim \frac{1}{c+\Phi^2}$.
This could be interpreted as a precursor of logarithmic tails 
expected in the Kondo regime \cite{tails}.

We used this 
function to fit the low-voltage MV-QMC data in Fig.~\ref{fig:vglSNRGSuscept}. The first thing to note is that
the low-voltage fit results in an excellent description of the SNRG data at high voltages, too. 
For $B=0.01\Gamma$, where the MV-QMC data have the largest error, a fit with all three parameters in 
Eq.\ \ref{eq:fitmagn} became very bad.  However, for larger $B$ with higher quality data, we observe that the 
parameter $c\approx2$.
Constraining this parameter to $c=2$ for $B=0.01\Gamma$ again results in an excellent fit here, too, including
the high-voltage data from SNRG. Further analyzing the field dependence of the parameters $a$ and $b$, 
we find a surprisingly simple behavior, viz $a\approx 0.75\cdot B/T_K$ and $b\approx2(1-B/T_K)$. We therefore
suggest as an approximate, but very accurate formula for the behavior of the local susceptibility  at low temperatures and fields $T,B\lesssim T_K$
\begin{equation}
\chi(T,B,\Phi) \approx \frac{\chi(T,B,0)}{\frac{{\tilde
\Phi}^2/4}{\sqrt{\left(1-B/T_K\right)^2 + {\tilde\Phi}^2/4}} + 1}.
\label{eq:fitmagn_full}
\end{equation}

\section{Summary}

In the present paper we have provided a comparison of three state-of-the-art computational
methods for two local observables on a quantum dot in a stationary nonequilibrium state, namely the double
occupancy and the magnetization. For the double occupancy, which only
indirectly exhibits the Kondo temperature as a relevant energy scale, we have found
substantial disagreement between all three methods, whose origin is unclear.
However, a qualitative agreement between RT-QMC and MV-QMC is found, 
and for the interaction parameters considered, the RT-QMC gives results in close agreement with second-order perturbation
theory.

With regard to the magnetization, we compared MV-QMC and
SNRG data within the Kondo regime. In contrast to the double occupancy, we
found good agreement between the two methods. This finding is remarkable, because
in these calculations, the physics is clearly controlled by the non-equilibrium Kondo effect.

\acknowledgments
We thank K.~Sch\"onhammer, M.~Jarrell, and A.~Schiller for useful discussions.  
This work was  supported by   the Deutsche
Forschungsgemeinschaft under Grant No.\ AN 275/6-2,
FP7/ERC starting grant No.~278023, and the NSF with Grant No.\ DMR-0907150.
We would like to acknowledge computer support from the HLRN,
the GoeGrid initiative, the GWDG,
and the NIC, Forschungszentrum J\"ulich,
under Project No.\ HHB000.
Parts of the implementation are based on the ALPS 1.3 library \cite{alps}.

\end{document}